\documentclass{cjpsuppl}
\usepackage{graphicx}
\begin{document}
\title{What we have learned from direct CP violation studies\\
       in kaon decays}
\authori{A.\,A. Belkov, A.\,V. Lanyov}
\addressi{Joint Institute for Nuclear Research, 
          Laboratory of Particle Physics,\\ 
          141980 Dubna, Moscow region, Russia}
\authorii{G. Bohm}
\addressii{DESY Zeuthen, Platanenallee 6, D-15735 Zeuthen, Germany}
\authoriii{}   \addressiii{}
\authoriv{}    \addressiv{}
\authorv{}     \addressv{}
\authorvi{}    \addressvi{}
\headtitle{What we have learned from direct CP violation \ldots}
\headauthor{A.A. Belkov, G.Bohm, and A.V. Lanyov}
\lastevenhead{A.A. Belkov: What we have learned from direct CP violation 
                           \ldots}
\refnum{}
\daterec{}
\suppl{A}  \year{2003} \setcounter{page}{1}
\maketitle

\begin{abstract}
    A self-consistent analysis of $K\to 2\pi$ and $K\to 3\pi$ decays within 
a unique framework of chiral dynamics applied to the QCD-corrected weak 
nonleptonic quark lagrangian has been performed.
    The results on $K\to 2\pi$ amplitudes at $O(p^6)$, including the value for
$\varepsilon^{'}/\varepsilon$, are compared with experiment to fix 
phenomenological $B$-factors for mesonic matrix elements of nonpenguin and
penguin four-quark operators. 
    The dependence of these $B$-factors on different theoretical uncertainties
and experimental errors of various input parameters is investigated.
    Finally, we present our estimates at $O(p^6)$ for the CP asymmetry of 
linear slope parameters in the $K^{\pm}\to 3\pi$ Dalitz plot.
\end{abstract}

\section{Introduction}

    The starting point for most calculations of nonleptonic kaon decays 
is an effective weak lagrangian of the form
\cite{vzsh,gilman-wise}
\begin{equation}
{\cal L}^{q}_{w}\big(|\Delta S|=1\big) =
\sqrt2\,G_F\,V_{ud}V^{*}_{us}\sum_{i} \widetilde{C}_i\,{\cal O}_i\,,
\label{weak-lagr}
\end{equation}
which can be derived with the help of the Wilson operator product
expansion (OPE) from elementary quark processes, with additional hard gluon
exchanges.
   In the framework of perturbative QCD the coefficients $\widetilde{C}_i$ 
are to be understood as scale and renormalization scheme dependent functions.
   There exist extensive next-to-leading order (NLO) calculations
\cite{buras1,ciuchini1} in the context of kaon decays, among others.
   These calculations are based on the possibility of factorization of
short- and long-distance contributions into Wilson coefficient
functions $C^{QCD}_i(\mu)$ and mesonic matrix elements of four-quark operators
${\cal O}_i$, respectively.
   The latter, however, can presently be obtained only by using
nonperturbative, i.e. model-dependent, methods, because not only perturbative
QCD breaks down at scales $\mu \le 1 \mbox{GeV}$, but also the QCD degrees of 
freedom (quarks and gluons) have to be replaced by the mesonic ones.

   Usually, the results of calculations are displayed with the help of
$B$-factors (bag parameters) in the form
\begin{equation}
T_{K\to 2\pi} = \sqrt2\,G_F\,V_{ud}V^{*}_{us} \sum_{i}
                 \Big[ C_i(\mu)B_i(\mu) \Big]
                 <\pi\pi |{\cal O}_i|K>_{vac.sat.}\,,
\label{B_definition}
\end{equation}
where the mesonic matrix elements of four-quark operators are approximated 
by their vacuum saturation values, which are real and $\mu$-independent.
   In principle, factors $B_i(\mu)$ should be estimated by some 
higher-order calculations in the long-distance regime, for instance, in
$1/N_c$-expansion \cite{buras2} in the form $1+O(1/N_c)$, or from the
lattice approach.
   The preliminary stage of these calculations is best characterized
by the long standing difficulties to explain quantitatively the well-known 
$\Delta I = 1/2$ rule.
   Recent lattice calculations \cite{pekurovsky} seem to succeed in this 
respect, but are at the same time at variance (even in sign) with
experimental values of $\mbox{Re}(\varepsilon^{'}/\varepsilon)$.
   Of course, the severe difficulties of all calculations of long-distance 
effects from ``first principles'' restricts the predictive power of 
(\ref{weak-lagr}), leaving only the possibility of some semi-phenomenological 
treatment \cite{buras1,buras3,buras4}, which cannot be used to test the 
Standard Model (SM) in the kaon sector in a fundamental way.

   The main aim of the present paper is a further semi-phenomenological
analysis of the long-distance (nonperturbative) aspects of the
above lagrangian, especially in view of the actuality of the task to
analyze the implications of the measured parameter of direct CP violation, 
$\varepsilon^{'}/\varepsilon$, on an alternative manifestation of direct CP 
violation by asymmetries in charged $K^\pm\to 3\pi$ decays.
   As the main result, we present numerical estimates of bag parameters and 
resulting asymmetries, in the form of probability densities, in order to 
demonstrate the experimental and theoretical uncertainties.

\section{General scheme of calculations}

  In (\ref{weak-lagr}), ${\cal O}_i$ are the four-quark operators, defined
either by combinations of products of quark currents ($i=1,2,3,4$, 
non-penguin diagrams) or, in case of gluonic ($i=5,6$) and electro-weak 
($i=7,8$) penguin operators, by products of quark densities.
   The operators ${\cal O}_i$ with $i=1,2,3,5,6$ describe weak transitions
with isospin change $\Delta I=1/2$ while the operator ${\cal O}_4$ corresponds
to a $\Delta I=3/2$ transition and operators ${\cal O}_{7,8}$ -- to mixture of 
$\Delta I=1/2$ and $\Delta I=3/2$ amplitudes.

   The operators ${\cal O}_i$ used here and in earlier work 
\cite{CP-enhancement,bos-weak,k2pi-our} differ from those used in 
\cite{buras1,buras2,buras3,buras4} and other papers 
\cite{ciuchini1,other_pap,dortmund,wu} (usually called $Q_i$).
   Both sets are related by linear relations, which are given for easy 
reference below:
\begin{eqnarray*}
&&Q_1=  2{\cal O}_1 +\frac{2}{5}{\cal O}_2 +\frac{4}{15}{\cal O}_3
       +\frac{4}{3}{\cal O}_4\,,\\
&&Q_2= -2{\cal O}_1 +\frac{2}{5}{\cal O}_2 +\frac{4}{15}{\cal O}_3
       +\frac{4}{3}{\cal O}_4\,,\\
&&Q_3=  2{\cal O}_1 + 2{\cal O}_2\,,\\
&&Q_4= -2{\cal O}_1 + 2{\cal O}_2\,,\\
&&Q_5=  4{\cal O}_6\,,\qquad Q_6= 2{\cal O}_5 + \frac{4}{3}{\cal O}_6\,,\\
&&Q_7=  {\cal O}_7\,,\qquad~~ Q_8= \frac{1}{2}{\cal O}_8 
                                  +\frac{1}{3}{\cal O}_7\,.
\end{eqnarray*}

   The general scheme of the meson matrix element calculation by using
the weak lagrangian (\ref{weak-lagr}) is based on the quark
bosonization approach \cite{bos-weak}. 
   The bosonization procedure establishes a correspondence between 
quark and meson currents (densities) and products of currents (densities).
   Finally, it leads to an effective lagrangian for nonleptonic kaon decays 
in terms of bosonized (meson) currents and densities:
$$
\bar q \gamma_\mu \frac14 (1\mp \gamma^5) \lambda^a q\,\,\, 
\Rightarrow\,\, J^{a\,(mes)}_{L/R \, \mu}\,, 
\quad
\bar q \frac14 (1\mp \gamma^5) \lambda^a q\,\,\,
\Rightarrow\,\, J^{a\,(mes)}_{L/R}\,.
$$

   The meson currents/densities $J^a_{L/R\mu}$ and $J^a_{L/R}$ are obtained 
from the quark determinant by variation over additional external sources
associated with the corresponding quark currents and densities \cite{bos-weak}.
   From the momentum expansion of the quark determinant to $O(p^{2n})$
one can derive the strong lagrangian for mesons ${\cal L}_{eff}$ of
the same order and the corresponding currents and densities $J^a_{L/R\mu}$ 
and $J^a_{L/R}$ to the order $O(p^{2n-1})$ and $O(p^{2n-2})$, respectively.
   Thus, the bosonization approach gives us the correspondence between
power counting for the momentum expansion of the effective chiral
lagrangian of strong meson interactions,
$$
{\cal L}^{(mes)}_s  =
{\cal L}_s^{(p^2)} + {\cal L}_s^{(p^4)} + {\cal L}_s^{(p^6)} + \,.\,.\,.\,,
$$ 
and power counting for the meson currents and densities:
\begin{eqnarray*} 
{\cal L}_s^{(p^n)}\,\,\Rightarrow\,\,J^{(p^{n-1})}_\mu\,\,\mbox{(currents)}\,;
\quad 
{\cal L}_s^{(p^n)}\,\,\Rightarrow\,\,J^{(p^{n-2})}\,\,\mbox{(densities)}\,.
\end{eqnarray*} 

   Some interesting observations on the difference of the momentum behavior
of penguin and non-penguin operators can be drawn from power-counting
arguments.
   The leading contributions to the vector currents and scalar densities are 
of $O(p^1)$ and $O(p^0)$, respectively.
   Since in our approach the non-penguin operators are constructed out of
the products of currents $J^a_{L\mu}$, while the penguin operators
are products of densities $J^a_L$, the lowest-order contributions
of non-penguin and penguin operators are of $O(p^2)$ and $O(p^0)$,
respectively.
   However, due to the well-known cancellation of the contribution of the
gluonic penguin operator ${\cal O}_5$ at the lowest order 
\cite{chivukula}, the leading gluonic penguin as well as non-penguin
contributions start from $O(p^2)$
\footnote{There is no cancellation of the contribution of the
          electromagnetic penguin operator ${\cal O}_8$ at the lowest
          order and the leading contributions start in this case from 
          $O(p^0)$}.
   Consequently, in order to derive the currents which contribute 
to the non-penguin transition operators at the leading order, it is sufficient 
to use the terms of the quark determinant to $O(p^2)$ only.
   At the same time the terms of the quark determinant to $O(p^4)$ have to 
be kept for calculating the penguin contribution at $O(p^2)$, since it
arises from the combination of densities, which are of $O(p^0)$ and 
$O(p^2)$, respectively.
   In this subtle way the difference in momentum behavior is revealed
between matrix elements for these two types of weak transition operators;
it manifests itself more drastically in higher-order lagrangians and
currents.

   This fact makes penguins especially sensitive to higher order effects.
   In particular, the difference in the momentum power counting behavior
between penguin and non-penguin contributions to the isotopic amplitudes
of $K\to 3\pi$ decays, which appears in higher orders of chiral theory,
leads to a dynamical enhancement of the charge asymmetry of the Dalitz-plot
linear slope parameter \cite{CP-enhancement,bos-weak}.

   In our approach the Wilson coefficients $\widetilde{C}_i$ in the effective
weak lagrangian (\ref{weak-lagr}) are treated as phenomenological parameters 
which should be fixed from experiment. 
   They are related with the Wilson coefficients $C^{QCD}_i(\mu)$,
calculated in perturbative QCD \cite{buras3}, via the 
$\widetilde{B}_i$-factors at the fixed renormalization scale $\mu =1$~GeV: 
$$
\widetilde{C}_i=C^{QCD}_i(\mu)\widetilde{B}_i(\mu)|_{\mu = 1GeV}\,.
$$
   The coefficients $C^{QCD}_i(\mu)$ can be written in the form of a sum of 
$z$ and $y$ components,
\begin{equation}
\widetilde{C}_i^{QCD}(\mu) = \widetilde{C}^{(z)}_i(\mu)
                             +\tau\,\widetilde{C}^{(y)}_i(\mu)\,,\quad
\tau = -\frac{V_{td}V_{ts}^{*}}{V_{ud}V_{us}^{*}}\,,
\label{wilson}
\end{equation}
where only the $y$ component contains the combination of CKM-matrix elements 
$\tau$, which, according to the SM, is responsible for CP violation -- it has
a non-vanishing imaginary part.
   For the amplitudes of nonleptonic kaon decays follows the same decomposition
into $z$- and $y$- components,
\begin{equation}
{\cal A}_I = {\cal A}^{(z)}_I+\tau{\cal A}^{(y)}_I\,,
\label{amplitude}
\end{equation}
where ${\cal A}_I$ are isotopic amplitudes of $K\to 2\pi$ decay corresponding 
to the $\pi\pi$ final states with the isospin $I=0,2$.

    With the dominating contributions ${\cal A}^{(i)}_I$ from the four-quark 
operators ${\cal O}_i$, the $K\to (2\pi)_I$ amplitudes may be written as
\begin{eqnarray*}
{\cal A}^{(z,y)}_I&=&
   \Big[-C_1^{(z,y)}(\mu)+C_2^{(z,y)}(\mu)+C_3^{(z,y)}(\mu)\Big]
    \widetilde{B}_1(\mu)\,{\cal A}^{(1)}_I
\\ &&
  +C_4^{(z,y)}(\mu)\widetilde{B}_4(\mu)\,{\cal A}^{(4)}_I
  +C_5^{(z,y)}(\mu)\widetilde{B}_5(\mu)\,{\cal A}^{(5)}_I
  +C_8^{(z,y)}(\mu)\widetilde{B}_8(\mu)\,{\cal A}^{(8)}_I\,.
\label{components}
\end{eqnarray*}
    In the calculation of the ${\cal A}^{(i)}_I$, sizable corrections connected
to isospin symmetry breaking have been taken into account (see below). 

    The observable effects of direct CP violation in the nonleptonic kaon 
decays are caused by the $y$-components of their amplitudes (\ref{amplitude}).
    In particular, the ratio $\varepsilon^{'}/\varepsilon$ can be expressed as
$$
\mbox{Re}\bigg( \frac{\varepsilon^{'}}{\varepsilon} \bigg) =
\mbox{Im}\,\lambda_t\,\big(P_0-P_2),\quad
P_I = \frac{\omega}{\sqrt{2}|\varepsilon| |V_{ud}||V_{us}|}\,
      \frac{{\cal A}_I^{(y)}}{{\cal A}_I^{(z)}}\,,
$$
where $\mbox{Im}\,\lambda_t = \mbox{Im}\,V^{*}_{ts}V_{td}
                            = |V_{ub}||V_{cb}| \mbox{sin} \delta$;
$\omega = {\cal A}_2^{(z)}/{\cal A}_0^{(z)}$.

\section{Theoretical and phenomenological uncertainties}

    There are the following theoretical and phenomenological uncertainties 
which appear both from short distance (Wilson coefficients) and long distance 
(effective chiral lagrangians and $\widetilde{B}_i$-factors) contributions to 
the kaon decay amplitude:
\begin{itemize}
\item Dependence on the parameter $\mbox{Im}\,\tau \sim \mbox{Im}\,\lambda_t$ 
      arising from the imaginary part of Wilson coefficient (\ref{wilson}).

\item Regularization scheme dependence which arises when calculating Wilson
      coefficients beyond the leading order (LO) of QCD in the next-to-leading
      orders: naive dimensional regularization (NDR), t'Hooft-Veltman 
      regularization (HV).

\item Dependence of Wilson coefficients on choice of the renormalization point
      $\mu$, taken below as 1~GeV, and QCD scale 
      $\Lambda^{(4)}_{\overline{MS}}$, contained in the interval 
      $(325\pm 110)$~MeV.

\item Factors $\widetilde{B}_i$ $(i=1,4,5,8)$ for dominating contributions of 
      four-quark operators ${\cal O}_i$ to the meson matrix element
      (\ref{components}).

\item Dependence of meson matrix elements on the structure constants $L_2$, 
      $L_3$, $L_4$, $L_5$, $L_8$ of the general form of the effective chiral
      lagrangian introduced at $O(p^4)$ by Gasser and Leutwyler \cite{gasser1}.

\item Dependence of meson matrix elements on the structure constants of the 
      effective chiral lagrangian at $O(p^6)$ \cite{p6-our}.

\item Dependence on choice of the regularization scheme to fix the UV 
      divergences resulting from meson loops.
\end{itemize}

    In the present paper we have combined a new systematic calculation of 
mesonic matrix elements for nonleptonic kaon decays from the effective chiral 
lagrangian approach with Wilson coefficients $C^{QCD}_i(\mu)$, derived by the 
Munich group \cite{buras3} for $\mu =1$ GeV and $m_t=170$ GeV.
    For the parameter $\mbox{Im}\,\lambda_t$ we have used the result
obtained in \cite{buras5}: $\mbox{Im}\,\lambda_t = (1.33 \pm
0.14)\cdot 10^{-4}$.
    We performed a complete calculation of $K\to 2\pi$ and $K\to 3\pi$
amplitudes at $O(p^6)$ including the tree level, one- and two-loop diagrams.
    For the structure coefficients $L_i$ of the effective chiral lagrangian at 
$O(p^4)$ we used the values fixed in \cite{dafne} from the phenomenological
analysis of low-energy meson processes.
    The structure coefficients of the effective chiral lagrangian at $O(p^6)$ 
have been fixed theoretically from the modulus of the logarithm of the quark 
determinant of the NJL-type model (see \cite{p6-our} for more details).
    The superpropagator regularization has been applied to fix UV divergences
in meson loops, thereby the renormalization scale was $\mu_{SP}=1$~GeV.
    The isotopic-symmetry-breaking corrections which arises from both the
$\pi^0$-$\eta$ and $\pi^0$-$\eta^{'}$ mixing and the final-state 
$\pi^\pm -\pi^0$ mass difference were taken into account
\footnote{In \cite{k2pi-our} the $\pi^0$-$\eta$-$\eta^{'}$ mixing
          contribution to the isospin breaking has been computed only at tree 
          level. The importance of the correction due to the final-state 
          $\pi^\pm -\pi^0$ mass difference was discussed for the first time 
          recently in \cite{suzuki}. This mass difference leads to a sizable 
          correction to $\varepsilon^{'}/\varepsilon$, in spite of its 
          smallness.}.
    In \cite{k2pi-our} one can find more technical details of the calculation
of $K\to 2\pi$ amplitudes.
  
\section{Phenomenological analysis of $K\to 2\pi$ decays}

    In our phenomenological analysis the results on the $K\to 2\pi$ 
amplitudes, including the value for $\varepsilon^{'}/\varepsilon$,
are compared with experiment to fix the phenomenological 
$\widetilde{B}_i$-factors for the mesonic matrix elements of nonpenguin and 
penguin four-quark operators. 
    As experimental input we used the experimental values of the isotopic 
amplitudes ${\cal A}_{0,2}^{(exp)}$ fixed from the widths of $K\to 2\pi$ 
decays and the world average value 
$\mbox{Re}\,\varepsilon^{'}/\varepsilon = (16.2\pm 1.7)\times 10^{-4}$
which includes both old results of NA31 \cite{NA31-CP} and E731 \cite{E731-CP}
experiments and recent results from KTeV \cite{KTeV-CP} and NA48 
\cite{NA48-CP}.
    The output parameters of the performed $K\to 2\pi$ analysis are the 
factors $\widetilde{B}_1$, $\widetilde{B}_4$ and $\widetilde{B}_5$ for
a fixed value of $\widetilde{B}_8$.

\begin{figure}
\begin{center}
\includegraphics[width=12cm]{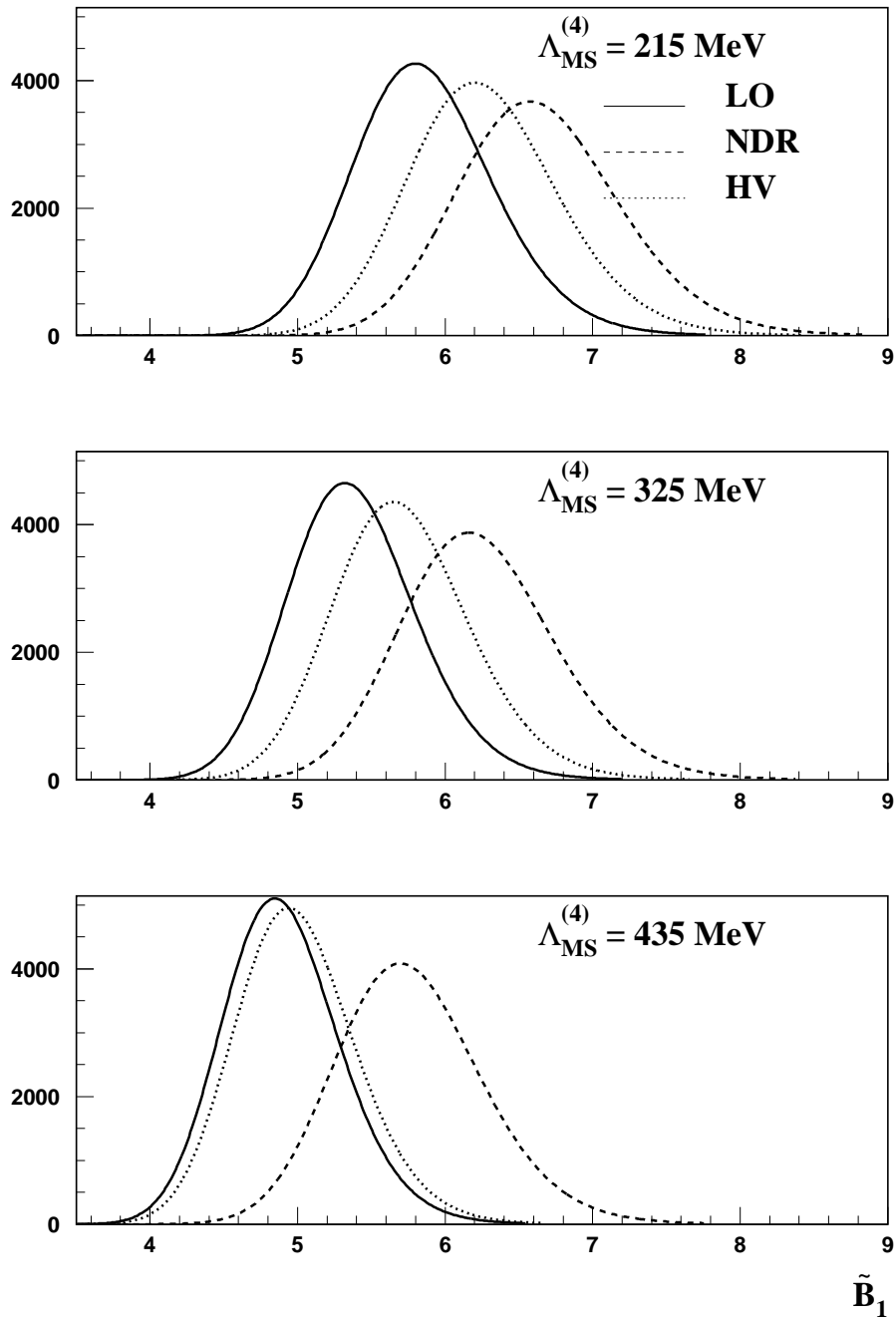}
\end{center}
\vspace*{-10mm}
\caption{Probability density distributions for factor $\widetilde{B}_1$ with 
         $\widetilde{B}_8=1$}
\label{b1}
\end{figure}

\begin{figure}
\begin{center}
\includegraphics[width=12cm]{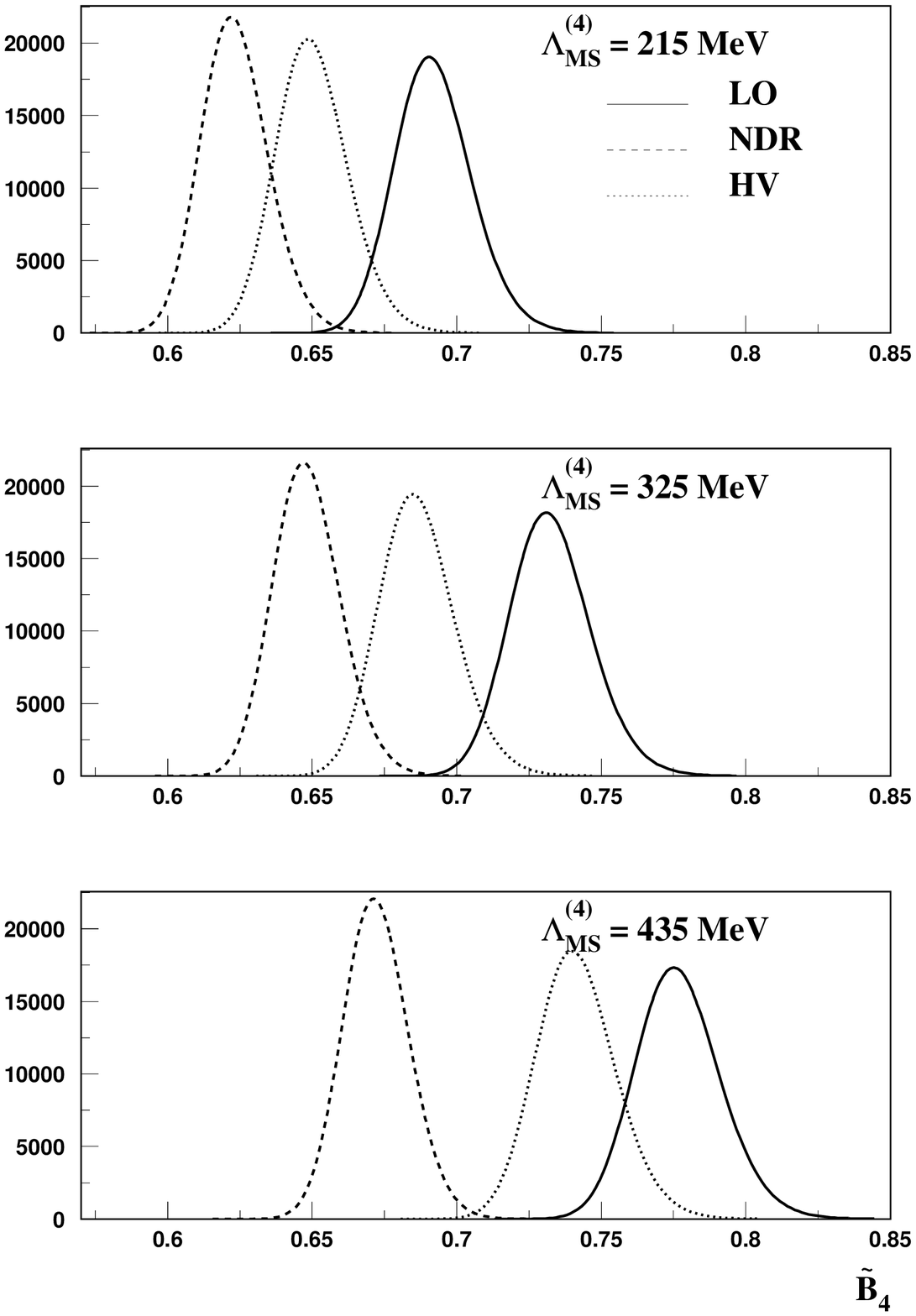}
\end{center}
\vspace*{-10mm}
\caption{Probability density distributions for factors $\widetilde{B}_4$ with 
         $\widetilde{B}_8=1$}
\label{b4}
\end{figure}

\begin{figure}
\begin{center}
\includegraphics[width=12cm]{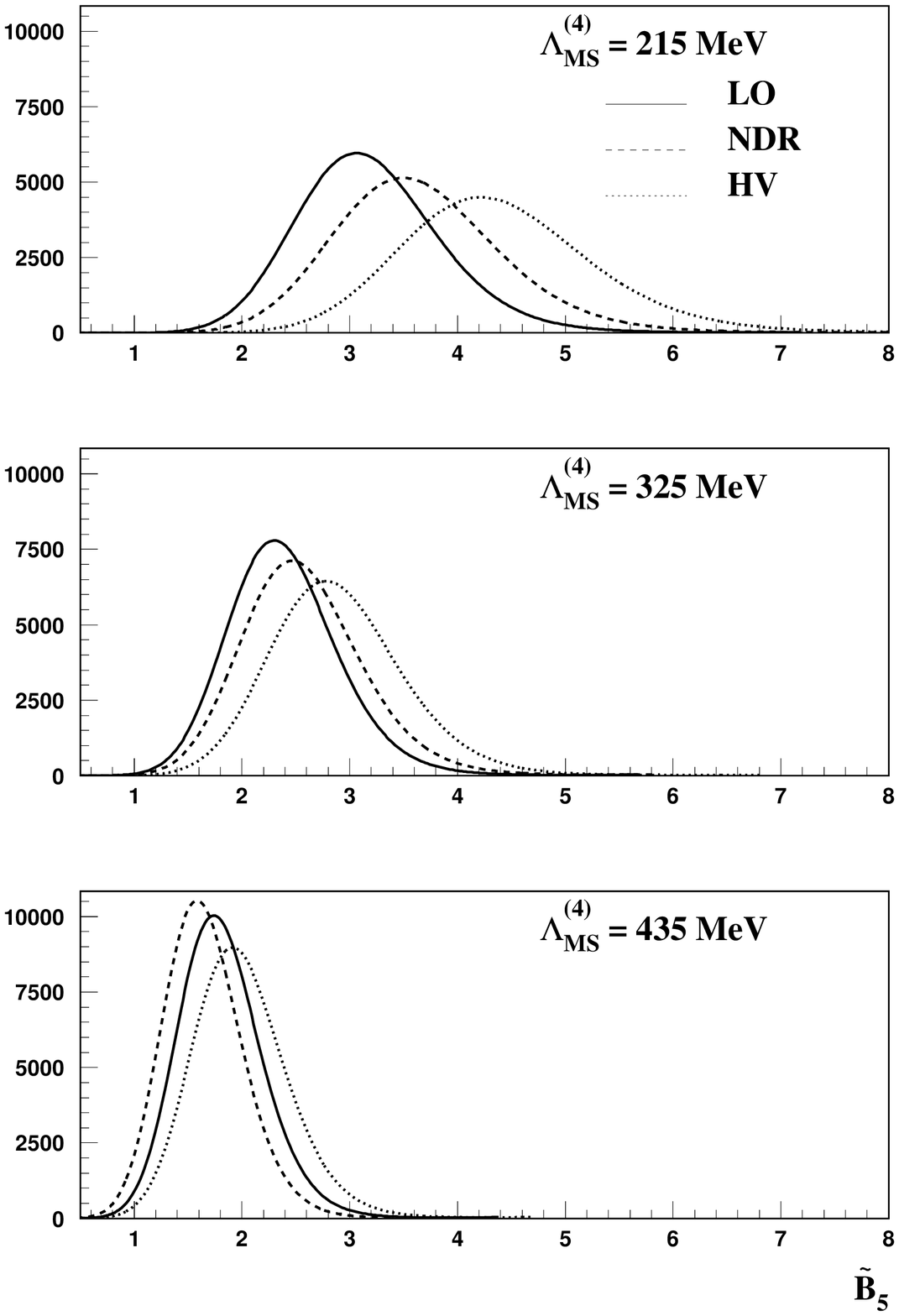}
\end{center}
\vspace*{-10mm}
\caption{Probability density distributions for factors $\widetilde{B}_5$ with 
         $\widetilde{B}_8=1$}
\label{b5}
\end{figure}

    The dependence of the $\widetilde{B}_i$-factors on different theoretical 
uncertainties and experimental errors of various input parameters is 
investigated by applying the ``Gaussian'' method. 
    Using  Wilson coefficients derived in \cite{buras3} in various
regularization schemes (LO, NDR, HV) for different values of the QCD scale 
$\Lambda^{(4)}_{\overline{MS}}$ ,we calculated the probability density 
distributions for $\widetilde{B}_i$-factors obtained by using Gaussian 
distributions for all input parameters with their errors.
    As an example, the probability densities for the parameters 
$\widetilde{B}_1$, $\widetilde{B}_4$, $\widetilde{B}_5$ calculated with
$\widetilde{B}_8=1$ and $\Lambda^{(4)}_{\overline{MS}}=325$ MeV are shown 
in Figs.~\ref{b1}, \ref{b4} and \ref{b5}.
    Upper and lower bounds for $B_i$-factors $(i=1,4,5)$ for different values 
of $\Lambda^{(4)}_{\overline{MS}}$ in LO, NDR and HV regularization schemes 
($\widetilde{B}_8 = 1$) obtained by the Gaussian method are shown in 
table~\ref{b_range}.
    The limits without parentheses correspond to the confidence level of 68\% 
while the limits in parentheses -- to the confidence level of 95\%.

    Fig.~\ref{b5} shows the necessity for a rather large gluonic penguin 
contribution to describe the recently confirmed significant experimental
$\varepsilon^{'}$ value (the factor $\widetilde{B}_5$ is found well above 1). 
    It should be emphasized that the dependence of $\widetilde{B}_i$ 
($i=1,4,5$) on $\widetilde{B}_8$ is very small even within a wide range of its
values $0 \le \widetilde{B}_8 \le 10$.
    Therefore even for $\widetilde{B}_8=0$ values of $\widetilde{B}_5 > 2$
are necessary to explain the large value of $\varepsilon^{'}/\varepsilon$.

\begin{figure}
\begin{center}
\includegraphics[width=12cm]{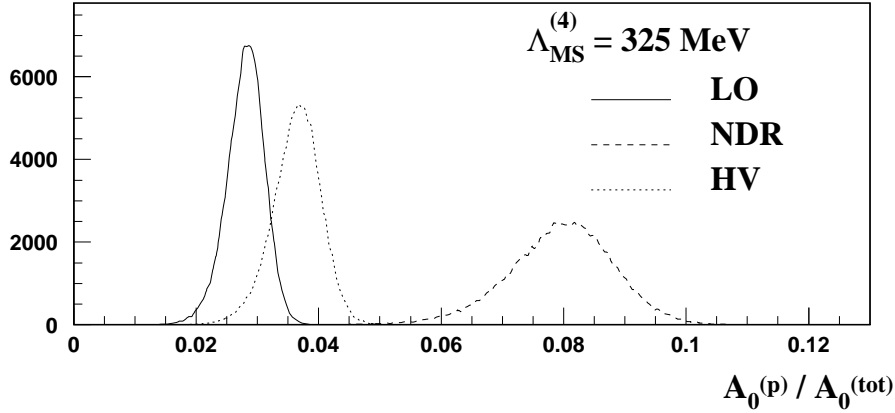}
\end{center}
\vspace*{-10mm}
\caption{Probability density distributions for the relative contribution 
         of penguin operators to the $\Delta I = 1/2$ amplitude}
\label{a0_rat-325.eps}
\end{figure}

    Even for larger values of $\widetilde{B}_5$, the contributions of 
nonpenguin operators to the $\Delta I = 1/2$ amplitude are still dominating 
(see Fig.~\ref{a0_rat-325.eps} and also upper and lower bounds for the 
relative contribution  of penguin operators to the $\Delta I = 1/2$ amplitude 
in table~\ref{A0p_range}).
    The large $\widetilde{B}_1$ and $\widetilde{B}_5$ values may be a hint 
that the long-distance contributions, especially to $\Delta I = 1/2$ 
amplitudes, are still not completely understood.
    An analogous conclusion has been drawn in \cite{buras5}, where also
possible effects from physics beyond the Standard Model are discussed.

\vspace{3mm}
\begin{table}
\caption{ Upper and lower bounds for $\widetilde{B}_i$ factors ($i=1,4,5$).
         The limits without parentheses correspond to the confidence level of 
         68\% while the limits in parentheses -- to the confidence level of 
         95\%.
}
\vspace{3mm}
\begin{center}
{\small
\begin{tabular}{|c|c|*{2}{c}|*2{c}|*2{c}|} \hline \hline
$~B_i~$& $\Lambda ^{(4)}_{\overline{MS}}$ & 
\multicolumn{2}{|c|}{LO}&\multicolumn{2}{|c|}{NDR}&\multicolumn{2}{|c|}{HV}\\
\cline{3-8}
$~~~~~$&(MeV)&  min &  max &  min &  max &  min &  max   \\
\hline
       & 215 &  5.4  & 6.4   &  6.1  & 7.2   &  5.8  & 6.8    \\
       &     &( 5.0  & 6.9  )&( 5.7  & 7.9  )&( 5.3  & 7.4  ) \\ \cline{2-8}
$~\widetilde{B}_1~$
       & 325 &  4.9  & 5.8   &  5.7  & 6.8   &  5.2  & 6.2    \\
       &     &( 4.6  & 6.3  )&( 5.3  & 7.4  )&( 4.7  & 6.7  ) \\ \cline{2-8}
       & 435 &  4.5  & 5.3   &  5.3  & 6.3   &  4.6  & 5.4    \\
       &     &( 4.2  & 5.8  )&( 4.9  & 6.8  )&( 4.3  & 5.9  ) \\ \hline \hline

       & 215 &  0.68 & 0.71  &  0.61 & 0.63  &  0.64 & 0.66   \\
       &     &( 0.67 & 0.72 )&( 0.60 & 0.65 )&( 0.62 & 0.68 ) \\ \cline{2-8}
$~\widetilde{B}_4~$
       & 325 &  0.72 & 0.75  &  0.64 & 0.66  &  0.67 & 0.70   \\ 
       &     &( 0.70 & 0.76 )&( 0.63 & 0.67 )&( 0.66 & 0.71 ) \\ \cline{2-8}
       & 435 &  0.76 & 0.79  &  0.66 & 0.68  &  0.73 & 0.76   \\
       &     &( 0.75 & 0.81 )&( 0.65 & 0.70 )&( 0.72 & 0.77 ) \\ \hline \hline

       & 215 &  2.5  & 3.9   &  2.8  & 4.5   &  3.5  & 5.3    \\
       &     &( 2.0  & 4.9 ) &( 2.3  & 5.6 ) &( 2.9  & 6.7  ) \\ \cline{2-8}
$~\widetilde{B}_5~$
       & 325 &  1.8  & 2.9   &  2.0  & 3.2   &  2.3  & 3.6    \\
       &     &( 1.5  & 3.7 ) &( 1.6  & 4.0 ) &( 1.8  & 4.5  ) \\ \cline{2-8}
       & 435 &  1.4  & 2.2   &  1.3  & 2.0   &  1.5  & 2.5    \\
       &     &( 1.1  & 2.9 ) &( 1.0  & 2.7 ) &( 1.2  & 3.1  ) \\ \hline \hline
\end{tabular}
}
\end{center}
\label{b_range}
\end{table}

\vspace{3mm}
\begin{table}
\caption{Upper and lower bounds for the relative contribution of penguin 
         operators to the $\Delta I = 1/2$ amplitude.
         The limits without parentheses correspond to the confidence level of 
         68\% while the limits in parentheses -- to the confidence level of 
         95\%.
}
\vspace{3mm}
\begin{center}
{\small
\begin{tabular}{|c|c|*{2}{c}|*2{c}|*2{c}|} \hline \hline
       & $\Lambda ^{(4)}_{\overline{MS}}$ & 
\multicolumn{2}{|c|}{LO}&\multicolumn{2}{|c|}{NDR}&\multicolumn{2}{|c|}{HV}\\
\cline{3-8}
$~~~~~$&(MeV)&  min &  max &  min &  max &  min &  max   \\
\hline
       & 215 & 0.023 & 0.029 & 0.062 & 0.078 & 0.031 & 0.039  \\
       &     &(0.019 & 0.031)&(0.053 & 0.084)&(0.027 & 0.042) \\ \cline{2-8}
$A^{(p)}_0/A^{(tot)}_0$
       & 325 & 0.025 & 0.031 & 0.071 & 0.088 & 0.032 & 0.040  \\
       &     &(0.021 & 0.034)&(0.061 & 0.096)&(0.027 & 0.044) \\ \cline{2-8}
       & 435 & 0.026 & 0.033 & 0.079 & 0.098 & 0.043 & 0.054  \\
       &     &(0.022 & 0.036)&(0.067 & 0.107)&(0.037 & 0.059) \\ \hline \hline
\end{tabular}
}
\end{center}
\label{A0p_range}
\end{table}

\section{Direct CP violation in $K\to 3\pi$ decays}

    Finally, the predictions for the CP asymmetry of linear 
slope parameters in the $K^{\pm}\to 3\pi$ Dalitz plot are discussed.
    These predictions are based on a new calculation of $K\to 3\pi$
amplitudes at $O(p^6)$ within the same effective lagrangian approach.
    The obtained  $K\to 3\pi$ amplitudes include the same theoretical
uncertainties as in case of the $K\to 2\pi$ analysis.  
    The values of $\widetilde{B}_1$, $\widetilde{B}_4$, $\widetilde{B}_5$
fixed from the $K\to 2\pi$ analysis are used as phenomenological input
to the $K\to 3\pi$ estimates which have been performed in a self-consistent
way by the Gaussian method.

    The linear slope parameter $g$ of the Dalitz plot for $K\to 3\pi$ decays 
is defined through the expansion of the decay probability in the kinematic
variables
$$
 |T(K\to 3\pi)|^2\,\,\propto\,\, 1 + gY +\,.\,.\,. 
$$
where $Y$ is a Dalitz variable:
$$
  Y = \frac{s_3-s_0}{m^2_\pi}\,,\quad
  s_3 = (p_K -p_{\pi_3})^2\,,\quad
  s_0 =  \frac{m^2_K}{3} + m^2_\pi\,,
$$
and the index $\pi_3$ belongs to the odd pion in the decays
$K^{\pm}\to \pi^{\pm}\pi^{\pm}\pi^{\mp}$ and 
$K^{\pm}\to \pi^{0}\pi^{0}\pi^{\pm}$.
    Direct CP violation leads to a charge asymmetry of the linear slope 
parameter,
$$
\Delta g(K^\pm \to 3\pi) = \frac{g(K^{+}\to 3\pi)-g(K^{-}\to 3\pi)}
                                {g(K^{+}\to 3\pi)+g(K^{-}\to 3\pi)}\,.
$$

\begin{figure}
\begin{center}
\includegraphics[width=12cm]{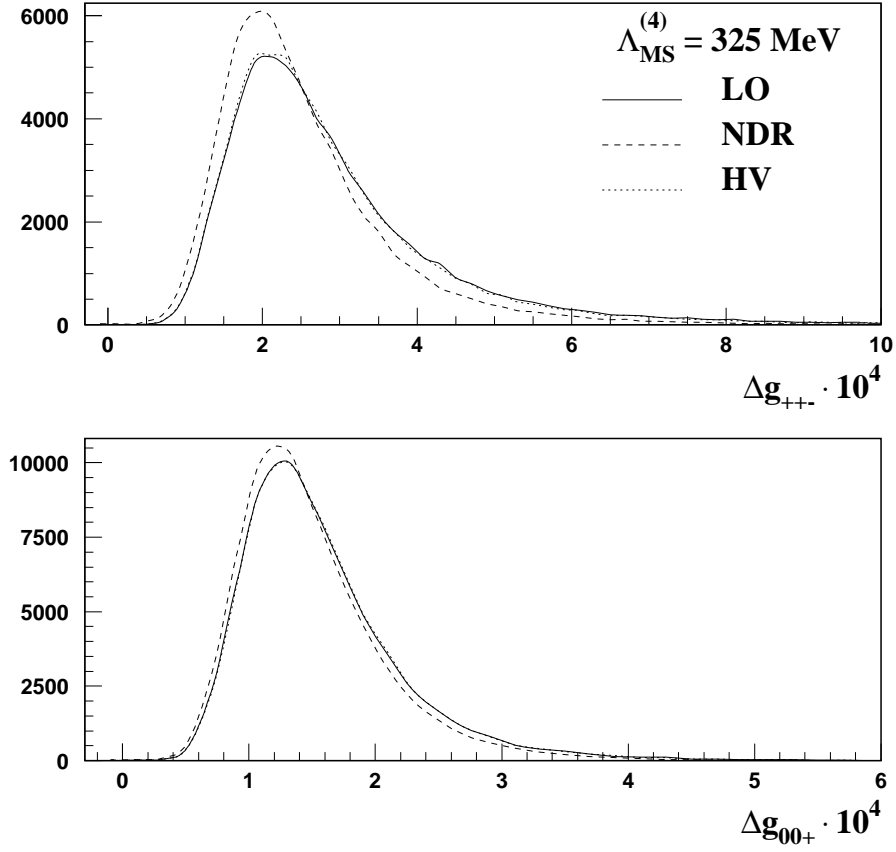}
\end{center}
\vspace*{-10mm}
\caption{Probability density distributions for the CP asymmetry of linear
         slope parameters of $K^{\pm}\to\pi^{\pm}\pi^{\pm}\pi^{\mp}$ and
         $K^{\pm}\to\pi^{0}\pi^{0}\pi^{\pm}$ decays}
\label{dg-325}
\end{figure}

    Fig.~\ref{dg-325} shows the probability density distributions for 
$K^{\pm}\to\pi^{\pm}\pi^{\pm}\pi^{\mp}$ and 
$K^{\pm}\to\pi^{0}\pi^{0}\pi^{\pm}$ decays calculated with
$\widetilde{B}_8=1$ and $\Lambda^{(4)}_{\overline{MS}}=325$ MeV.
    Upper and low bounds for $\Delta g_{++-}$ and $\Delta g_{00+}$ for 
different values of $\Lambda^{(4)}_{\overline{MS}}$ in LO, NDR and HV
regularization schemes ($\widetilde{B}_8 = 1$) obtained by the Gaussian method 
are shown in table~\ref{dg_range}.
    Summarizing these results, we have obtained the following upper and lower 
bounds for the charge symmetries of the linear slope parameter:
$$
      1.6\, <\,\Delta g_{++-}\cdot 10^4\,<\, \,4.2\,,\quad
      0.9\, <\,\Delta g_{00+}\cdot 10^4\,<\, \,2.2\,\,\,\, 
      \mbox{with\,\,CL=68\%};
$$
$$ 
      1.1\, <\,\Delta g_{++-}\cdot 10^4\,< \,7.6\,,\quad
      0.6\, <\,\Delta g_{00+}\cdot 10^4\,< \,3.3\,\,\,\,
      \mbox{with\,\,CL=95\%}.
$$

\vspace{3mm}
\begin{table}
\caption{ Upper and low bounds for $\Delta g_{++-}$ and $\Delta g_{00+}$ 
         (in units $10^{-4}$).
         The limits without parentheses correspond to the confidence level of 
         68\% while the limits in parentheses -- to the confidence level of 
         95\%.
}
\vspace{3mm}
\begin{center}
{\small
\begin{tabular}{|c|c|*{2}{c}|*2{c}|*2{c}|} \hline \hline
$~\Delta g$& $\Lambda ^{(4)}_{\overline{MS}}$ & 
\multicolumn{2}{|c|}{LO}&\multicolumn{2}{|c|}{NDR}&\multicolumn{2}{|c|}{HV}\\
\cline{3-8}
$~~~~~$&(MeV)&  min &  max &  min &  max &  min &  max   \\
\hline
       & 215 &  1.6 & 4.1  &  1.5 & 3.6  &  1.6 & 4.1    \\
       &     &( 1.1 & 7.6 )&( 1.0 & 6.0 )&( 1.1 & 7.4 )  \\ \cline{2-8}
$\Delta g_{++-}$ 
       & 325 &  1.6 & 4.2  &  1.5 & 3.6  &  1.6 & 4.1    \\
       &     &( 1.1 & 7.6 )&( 1.0 & 5.9 )&( 1.1 & 7.3  ) \\ \cline{2-8}
       & 435 &  1.6 & 4.2  &  1.5 & 3.5  &  1.6 & 4.0    \\
       &     &( 1.1 & 7.6 )&( 1.0 & 5.7) &( 1.1 & 7.0  ) \\ \hline \hline
       & 215 &  0.9 & 2.2  &  0.9 & 2.1  &  0.9 & 2.2    \\
       &     &( 0.6 & 3.3 )&( 0.6 & 3.1 )&( 0.6 & 3.3 )  \\ \cline{2-8}
$\Delta g_{00+}$ 
       & 325 &  0.9 & 2.2  &  0.9 & 2.1  &  0.9 & 2.2    \\ 
       &     &( 0.6 & 3.4 )&( 0.6 & 3.1 )&( 0.6 & 3.3 )  \\ \cline{2-8}
       & 435 &  0.9 & 2.2  &  0.9 & 2.1  &  0.9 & 2.2    \\
       &     &( 0.6 & 3.4 )&( 0.6 & 3.0 )&( 0.6 & 3.3 )  \\ \hline \hline
\end{tabular}
}
\end{center}
\label{dg_range}
\end{table}

   When comparing these results with those found above for the phenomenological
$\widetilde{B}$-factors, one finds a much reduced dependence on the 
regularization scheme and on the scale $\Lambda ^{(4)}_{\overline{MS}}$.
   Once more this type of asymmetry ratios turns out to be more stable not 
only against systematic experimental errors (efficiencies, among others), but 
also with respect to theoretical uncertainties of its parts.

\section{Conclusion}

    With some experimental updates and theoretical refinements our new 
estimates confirm the dynamical enhancement mechanism for the charge asymmetry
$\Delta g$ by higher order contributions in the effective chiral lagrangian 
approach which was first observed in \cite{CP-enhancement}.
    The predicted slope parameter asymmetry, although small, may be in reach
of current high statistics experiments \cite{app}.
    Already with lower statistics, new measurements of quadratic slope 
parameters of $K\to 3\pi$ decays, including the neutral channels, would lead 
to an improved theoretical understanding of the nonperturbative part of 
nonleptonic kaon decay dynamics.

   For $\varepsilon^{'}/\varepsilon$, the results of analysis are very 
sensitive to final-state interactions, isotopic-symmetry-breaking effects and 
other refinements of the model calculations.
   Despite of considerable theoretical efforts dedicated to the calculation of
Wilson coefficients and bag factors, the remaining uncertainties are still very
large (compare also \cite{hambye}, where a similar conclusion is reached,
when taking into account the full range of all input parameters and all
theoretical uncertainties).
   The fact that phenomenological values of $\widetilde{B}_i$-factors 
($i=1,4,5$) considerably differ from unity shows that the long-distance 
contributions are still not completely understood.
   Besides the (rather remote) possibility, that the SM has to be revised in
the kaon sector, there are further, more exotic developments in 
long-distance QCD going on \cite{kochelev}, which may have manifestations 
also here.  Another very active direction is the investigation of the problem 
of matching between short- and long-range renormalzation schemes by means of 
(effective) color singlet boson models. For a recent study, especially of CP 
violation in $K\to 3\pi$ decays, see \cite{prades} and papers cited there.

\bigskip


\end{document}